\documentstyle[aps,preprint]{revtex}
\begin{document}
\newcommand{\pp}[1]{\phantom{#1}}
\newcommand{\be}{\begin{eqnarray}}
\newcommand{\ee}{\end{eqnarray}}
\newcommand{\ve}{\varepsilon}
\newcommand{\Tr}{{\rm Tr\,}}
\newtheorem{th}{Theorem}
\newtheorem{lem}[th]{Lemma}

\title{
Manifestly Covariant Approach to Bargmann-Wigner Fields (I):\\
Generalized scalar products and Wigner states
}
\author{Marek Czachor \cite{*}}
\address{
Wydzia{\l}  Fizyki Technicznej i Matematyki Stosowanej\\
 Politechnika Gda\'{n}ska,
ul. Narutowicza 11/12, 80-952 Gda\'{n}sk, Poland
}
\maketitle
\begin{abstract}
Manifestly covariant formalism for 
Bargmann-Wigner fields is developed. 
 It is shown that there exists some
freedom in the choice of the form of the Bargmann-Wigner scalar
product: The general product depends implicitly on a family of
world-vectors. The standard choice of the product corresponds to
timelike and equal vectors which define a ``time" direction. 
The generalized form shows that formulas are simpler if one
chooses {\it null\/} directions. 
This freedom is used to derive
simple covariant formulas
for momentum-space wave functions (generalized Wigner states)
corresponding to arbitrary mass and spin and using eigenstates of the
Pauli-Lubanski vector. The eigenstates which
make formulas the simplest correspond to projections of the
Pauli-Lubanski vector on {\it null\/} directions. The new formulation
is an alternative to the standard helicity formalism. 
\end{abstract}

\section{Introduction}

The main objective of this paper is to 
develop a manifestly covariant formalism for 
Bargmann-Wigner fields\cite{BW}.
There are several
reasons for undertaking this task. One of them is to fill a sort
of gap between the powerful covariant
 spinor methods \cite{PR,PR2} and the noncovariant
methods of  induced representations of the
Poincar\'e group \cite{BR}. The noncovariance 
of the induced representations, and
especially of their generators, 
is manifested in  a particular decomposition of 
spinor generators into boosts and rotations which is used 
to represent boosts by the so-called
Wigner rotations \cite{Ohnuki}. In addition, a
transition between Wigner and spinor bases  involves
dividing a Fourier transform of the spinor field by some powers of
energy $p^0$. Typically, this $p^0$ is identified with the $p^0$
appearing in the invariant measure $d^3p/(2|p^0|)$ and  leads to
the characteristic additional powers $|p^0|^n$ appearing in the
spinor versions of the Bargmann-Wigner products. In the
covariant form of the scalar product given below these
additional powers of energy will be shown to possess some
arbitrariness which is normally hidden behind the noncovariance
of the standard expressions for the scalar products. 

The mentioned decomposition into boosts and rotations leads to
interpretational difficulties that are deeper than just the
lack of manifest covariance of formulas. A good example is
the celebrated problem of relativistic position operators
\cite{Pryce,Fleming,Bacry,JJ,K90}.
The operator which was shown to be a natural center-of-mass
position operator is 
$
\bbox Q=\frac{1}{2}\Bigl(H^{-1}\bbox K + \bbox K H^{-1}
\Bigr) 
$
where $H$ is a free Hamiltonian and $\bbox K$ generates boosts.
The position operator so obtained has many reasonable properties
and is, up to some domain questions, unique for massless fields 
\cite{JJ}. Still, the operator is obviously non-covariant since
no ``center-of-mass time operator" is known. 
The question of covariance was raised many years ago by Fleming
\cite{Fleming}, who proposed a hyperplane-dependent formalism.
The idea of hyperplane-dependent and Poincar\'e covariant
position operator appears, in a different context and
formulation, also in the unpublished thesis of Kaiser
\cite{thesis}. 
The author of the present work agrees generally
with Fleming's and Kaiser's 
conclusions that the hyperplane formalism is
natural for fields defined on Minkowski space. 
He believes, however, that
some of the interpretational difficulties have their origin in the
noncovariant formulation of the momentum-space wave functions.  

The methods developed below lead to manifestly covariant
formulas for the momentum-space wave functions in both massive
and massless cases, and for any spin. The formalism we introduce
naturally selects wave functions corresponding to 
projections of Pauli-Lubanski spin in
{\it null\/} directions. 
The same directions play a privileged
role in defining Bargmann-Wigner scalar products \cite{MC2}. 
The choice of
null directions differs from the standard choice of {\it
timelike\/} directions leading to the helicity formalism. The
latter has been extesively 
discussed in the 
context of quantum electrodynamics
and quantum optics 
\cite{moses1,moses2,moses3,ibb1,ibb2,ibb3,ibb4,sipe}.
Some formulas appearing in this paper look
surprisingly similar to those arising in 
Penrose's twistor formalism 
(\cite{PR2}, see also \cite{B1,B2}) even though
there exist essential mathematical differences between the two
formulations.  

Finally, to close the introductory remarks, let us note that 
Hilbert space methods related to the
Bargmann-Wigner scalar products
were shown recently to play an important role in
a wavelet formulation of 
electrodynamics \cite{K} (cf. also earlier results on
analytic-signal transform for massive fields 
\cite{jmp1,jmp2,jmp3,K90}). 
The results presented in this paper will prove
useful for a wavelet formulation of other Bargmann-Wigner fields
\cite{MC}.

The paper is organized as follows. In Sec.~\ref{M0} we discuss
massless fields and derive
new formulas for the Bargmann-Wigner norms. 
We show that the {\it form\/}
of the
generalized norms depends on some arbitrary world-vectors 
even though their {\it value\/} is independent of them. 
This freedom leads to a concrete form of Wigner states which
{\it do\/} depend on these vectors. 
In Sec.~\ref{Mn0} analogous formulas are derived for massive
fields. In several
appendices we explain the bispinor convention
which is used here, discuss properties of the Pauli-Lubanski
(P-L) vector and expansions of solutions of the Bargmann-Wigner
massive equations in terms of eigenstates of the P-L vector.
The new form of the Wigner states will lead to a new form of
generators which will be discussed in a forthcoming paper. 
The new generators will prove useful in discussing the question
of covariance of the relativistic center-of-mass position operator.

\section{Massless Bargmann-Wigner Fields}
\label{M0}

A massless spin-$n/2$ field is described by the spinor
equations \cite{PR}
\begin{eqnarray}
\nabla{^{A_k}}{_{A_k'}}\psi(x)_{A_1\dots 
A_k\dots A_rA'_1\dots A'_{r\pm n}}&=&0,\label{1}\\
\nabla{_{A_k}}{^{A_k'}}\psi(x)_{A_1\dots A_rA'_1\dots 
A'_k\dots A'_{r\pm n}}&=&0\label{2},
\ee
where the spinor $\psi(x)_{A_1\dots A_rA'_1\dots A'_{r\pm n}}$
is totally symmetric in all indices. For simplicity of notation
let us consider the case $r=n$, and a field which has only
unprimed indices. 

With any massless field one can associate locally 
various types of
potentials \cite{PR2}. 
The Hertz-type potentials are defined by
\be
\psi(x)_{A_1\dots  A_n}=\nabla_{A_1A'_1}\dots \nabla_{A_nA'_n}
\xi(x)^{A'_1\dots  A'_n}
\ee
with the subsidiary condition 
\be
\Box \xi(x)^{A'_1\dots  A'_n}=0.
\ee
Potentials of another type  are defined by
\be
\psi(x)_{A_1\dots  A_n}=\nabla_{A_1A'_1}\dots \nabla_{A_kA'_k}
\phi(x)^{A'_1\dots  A'_k}_{A_{k+1}\dots A_n},
\ee
and are subject to
\be
\nabla^{A_{k+1}A'_{k+1}}
\phi(x)^{A'_1\dots  A'_k}_{A_{k+1}\dots A_n}=0
\ee
implying the generalized Lorenz gauge
\be
\nabla{^{A_{k+1}}}{_{A'_{k}}}
\phi(x)^{A'_1\dots  A'_k}_{A_{k+1}\dots A_n}=0.
\ee
Let us begin with the Fourier representation of both
 the spinor field
and its Hertz-type potential:
\be
\psi(x)_{A_1\dots A_n}&=&
\frac{1}{(2\pi)^3}\int\frac{d^3p}{2p^0}\,
\Bigl\{
e^{-ip\cdot x}
\psi_+(p)_{A_1\dots A_n}
+
e^{ip\cdot x}
\psi_-(-p)_{A_1\dots A_n}
\Bigr\}\\
\xi(x)^{A'_1\dots A'_n}&=&
\frac{1}{(2\pi)^3}\int\frac{d^3p}{2p^0}\,
\Bigl\{
e^{-ip\cdot x}
\xi_+(p)^{A'_1\dots A'_n}
+
e^{ip\cdot x}
\xi_-(-p)^{A'_1\dots A'_n}
\Bigr\},
\ee
where $p$ is {\it future-pointing\/}. 
These definitions imply that
\be
\psi_\pm(\pm p)_{A_1\dots A_n}=(\mp i)^n 
p_{A_1A'_1}\dots p_{A_nA'_n}
\xi_\pm(\pm p)^{A'_1\dots A'_n}.\label{pH}
\ee
Define the tensor
\be
T_\pm(\pm p)_{a_1\dots a_n}&=&
\psi_\pm(\pm p)_{A_1\dots A_n}
\bar \psi_\pm(\pm p)_{A'_1\dots A'_n}\nonumber\\
&=&
p_{A_1B'_1}p_{ B_1A'_1}\dots p_{ A_nB'_n}p_{ B_nA'_n}
\xi_\pm(\pm p)^{B'_1\dots B'_n}
\bar \xi_\pm(\pm p)^{B_1\dots B_n}\nonumber\\
&=&
p_{ a_1}\dots p_{ a_n}p_{ b_1}\dots p_{ b_n}
U_\pm(\pm p)^{b_1\dots b_n},\label{tensor}
\ee
where we have used the trace-reversal spinor formula \cite{PR}
\be
p{_{A}}{_{B'}}p{_{B}}{_{A'}}=
p_{a}p_{b}-\frac{m^2}{2}g{_{a}}{_{b}}
\ee
and
\be
U_\pm(\pm p)^{b_1\dots b_n}=
\xi_\pm(\pm p)^{B'_1\dots B'_n}
\bar \xi_\pm(\pm p)^{B_1\dots B_n}.
\ee
The Poincar\'e (i.e. spinor) transformation of the 
field implies
\be
T'_\pm(\pm p)_{a_1\dots a_n}&=&
p_{ a_1}\dots p_{ a_n}p^{b_1}\dots p^{b_n}
U'_\pm(\pm p)_{b_1\dots b_n}\noindent\\
&=&
p_{ a_1}\dots p_{ a_n}p^{b_1}\dots p^{b_n}
\Lambda{_{b_1}}{^{c_1}}\dots\Lambda{_{b_n}}{^{c_n}}
U_\pm(\pm {\Lambda^{-1}p})_{c_1\dots c_n}\noindent\\
&=&
p_{ a_1}\dots p_{ a_n}
(\Lambda^{-1}p)^{b_1}
\dots(\Lambda^{-1}p)^{b_n}
U_\pm(\pm {\Lambda^{-1}p})_{b_1\dots b_n}.
\ee
Define 
\be
\parallel \psi_\pm\parallel'^2=
\int\frac{d^3p}{2p^0}
\frac{
t_1^{a_1}\dots t_n^{a_n}
T_\pm(\pm p)_{a_1\dots a_n}}{
t_1^{b_1}\dots t_n^{b_n}p_{ b_1}\dots p_{ b_n}}=
\int\frac{d^3p}{2p^0}
p^{a_1}\dots p^{a_n}
U_\pm(\pm p)_{a_1\dots a_n}\label{exx}
\ee
which is manifestly invariant and independent of the choice of 
$t_1^{a},\dots,t_n^{a}$ because the RHS of the equation 
does not depend on them. 
Consider $t_k^{a}=t^{a}$ 
where $t^{a}p_{ a}$ is equal to $p^0$ used in the
invariant measure.  

The matrix form of $t^{AA'}$
is (cf. \cite{PR})
\be
t^{{\bf AA}'}=t^{a}g{_a}{^{{\bf AA}'}}=
t^{0}g{_0}{^{{\bf AA}'}}=
\frac{1}{\sqrt{2}}
\left(
\begin{array}{cc}
1 & 0 \\
0 & 1
\end{array}
\right)
\ee
We have therefore
\be
\parallel \psi_\pm\parallel'^2=
\int\frac{d^3p}{2|p^0|^{n+1}}
T_\pm(\pm p)_{0\dots 0}.\label{scalK}
\ee
The Maxwell field corresponds to $n=2$. The resulting norm is
equivalent to this used by Kaiser in his construction of
electromagnetic wavelets \cite{K} (compare also Appendix
\ref{A3}). 

The formula~(\ref{exx}) can be generalized 
by  explicitly writing the ``time" direction $t^a$ in the
measure. Let $t\cdot t=1$. Define 
\be
t_a{^b}&=&g_a{^b}-t_at^b\\
{^t}p^a&=&t^a{_b}p^b\\
^tp&=&t^ap_a.
\ee
For general (massive or massless) $p^a$ we have 
\be
^tp^2=-t^{ab}p_ap_b + m^2=-\,^tp_a{^tp^a} + m^2.
\ee
For both $p^a$ and $t^a$ future-pointing we can define
\be
^t\psi_\pm(\pm ^tp^a)_{A_1\dots A_n}=
\psi_\pm\Bigl( \pm( 
{^t}p\,t^a + ^tp^a)\Bigr)
_{A_1\dots A_n},
\ee
and
\be
d\mu_{t,m}(^tp^a)
=\frac{e_{abcd}t^ad{\,^t}p^b\wedge d{\,^t}p^c \wedge d{\,^t}p^d}
{2\,^tp}.
\ee
Now
\be
\parallel \psi_\pm\parallel'^2=
\int d\mu_{t,0}(^tp^a)
\frac{
t_1^{a_1}\dots t_n^{a_n}
{\,^t}T_\pm(\pm ^tp^a)_{a_1\dots a_n}}{
t_1^{b_1}\dots t_n^{b_n}p_{ b_1}\dots p_{ b_n}}=
\int d\mu_{t,0}(^tp^a)
p^{a_1}\dots p^{a_n}
{\,^t}U_\pm(\pm ^tp^a)_{a_1\dots a_n}\label{gexx}
\ee
where ${\,^t}U_{\dots}$ and ${\,^t}T_{\dots}$ are defined in terms of
${\,^t}\psi_{\dots}$. The form (\ref{gexx}), which may seem artificial
and somewhat trivial, will prove very useful in analysis of
generators of the Poincar\'e group and generalized
analytic-signal transform \cite{K} for arbitrary 
Bargmann-Wigner fields \cite{MC}. 

The expression (\ref{exx}) is directly related to the
Bargmann-Wigner norm.

The well known fact that the field 
$
\psi(x)_{A_1\dots A_rA'_1\dots A'_{r\pm n}}
$
carries only one helicity can be shown in a covariant manner as
follows. We first contract the field equation (\ref{1})
with $g_a{^{B_kA'_k}}$ and use the identity (\ref{id1}) 
from Appendix~\ref{A1}. Performing an analogous transformation
of (\ref{2}) and
using the Pauli-Lubanski (P-L) vector (cf. Appendix~\ref{APL})
we obtain the equivalent forms of (\ref{1}) and (\ref{2})
\be
-\frac{1}{2}P^a\psi(x)_{A_1\dots 
A_rA'_1\dots A'_{r\pm n}}&=&S^a{_{A_k}}{^{B_k}}
\psi(x)_{A_1\dots B_k\dots
A_rA'_1\dots A'_{r\pm n}},\label{1'}\\
\frac{1}{2}P^a\psi(x)_{A_1\dots 
A_rA'_1\dots A'_{r\pm n}}&=&S^a{_{A'_k}}{^{B'_k}}
\psi(x)_{A_1\dots
A_rA'_1\dots B'_k\dots A'_{r\pm n}}.\label{2'}
\ee
We can further simplify the equations by
introducing the generators 
${\sigma}^{ab\pp A {\cal B}}_{\pp {aa}{\cal A}}$ 
of the $(r/2, r/2\pm n/2)$ representation. With the help of the 
respective
P-L vector the massless equation reduces to  
\be
\pm\frac{n}{2}P^a\psi(x)_{\cal A}&=&S^a{_{\cal A}}{^{\cal B}}
\psi(x)_{\cal B},\label{r,s}
\ee
where $\cal A$, $\cal B$  stand for $A_1\dots 
A_rA'_1\dots A'_{r\pm n}$, etc.

The eigenequation 
(\ref{r,s}) determines the Fourier components of the field up to
a $ p$-dependent factor (an ``amplitude"). 
We can write, therefore,
\be
\psi_\pm(\pm p)_{A_1\dots A_n}=(\mp i)^n 
p_{ A_1A'_1}\dots p_{ A_nA'_n}
\eta_{\pm}(\pm p)^{A'_1\dots A'_n}\,f_\pm(\pm p),\label{pH'}
\ee
where the only restriction on $f_\pm( p)$ is the
square-integrability of the field, and 
$\eta_{\pm}( p)^{A'_1\dots A'_n}$ is normalized by
\be
p_{ b_1}\dots p_{ b_n}
\eta_{\pm}(\pm p)^{B'_1\dots B'_n}
\bar \eta_{\pm}(\pm p)^{B_1\dots B_n}=1.
\ee
With this normalization (\ref{tensor}) becomes 
\be
T_\pm(\pm p)_{a_1\dots a_n}&=&
p_{ a_1}\dots p_{ a_n}
|f_\pm(\pm p)|^2.
\ee
We can choose $\eta_{\pm}(\pm p)^{A'_1\dots A'_n}$ as follows.
Let $p_{ a}= \pi_{ A}\,\bar \pi_{ A'}$, 
and let $\omega^{A}$ 
satisfy $\pi_{ A}\omega^{A}=1$ 
(i.e. the pair $\pi_{ A}$, $\omega^{A}$
is a spin-frame \cite{PR}). Then
\be
\eta_{\pm}(\pm p)^{A'_1\dots A'_n}=(\pm i)^n\bar \omega^{A'_1}\dots
\bar \omega^{A'_n},
\ee
and
\be
\psi_\pm(\pm p)_{A_1\dots A_n}=
\pi_{ A_1}\dots \pi_{ A_n}f_\pm(\pm p).\label{twist}
\ee
The amplitude satisfies 
\be
\parallel\psi_\pm\parallel'^2=
\int\frac{d^3p}{2p^0}\,
|f_\pm(\pm p)|^2=\parallel\psi_\pm\parallel^2.\label{bb1}
\ee
where $\parallel\psi_\pm\parallel^2$ is the Bargmann-Wigner
norm. 
Therefore $f_\pm(\pm p)$ is the Bargmann-Wigner 
amplitude which is used in \cite{ibb1,ibb2,ibb3,ibb4} in
the context of the electromagnetic field and the photon wave function.
The form (\ref{twist}) resembles kernels of contour integral
expressions for massless fields arising in the twistor formalism
(cf. \cite{PR2}, Eq.~(6.10.3)). 
The temptation to {\it identify\/} the Bargmann-Wigner amplitude with
the twistor wave function must be resisted despite the
apparent similarity of the two objects. The difference lies
essentially in the meaning of the integral. Here it is a
3-dimensional integral over the light cone (which could be
expressed directly in terms of spinors $\pi_A$ and $\bar
\pi_{A'}$) whereas in the twistor formalism we have complex
(1- or 2-dimensional) 
contour integrals in the space of spinors. Moreover the Fourier
transform contains the factors 
$
\exp (\pm i\pi_A\bar
\pi_{A'}x^{AA'})=\exp (\pm ip_ax^{a})
$
which are not holomorphic functions of either $\bar
\pi_{A'}$ or $\pi_{A}$, whereas the twistor formulation requires
wave functions which are holomorphic functions of twistors. 
Still the analogy to the twistor formalism could be pursued
further and, for example, a formula for the ``twistor-like" 
helicity operator could be directly derived. 

Instead of following in this direction
let us discuss briefly the freedom in the
choice of the Wigner states which is implied by the freedom in
the choice of the norm (\ref{exx}). 

Following the standard approaches (cf. \cite{Ohnuki}) we can
define Wigner states by
\be
\psi_\pm(\pm p,\{t_k\})_{A_1\dots A_n}=
\frac{\psi_\pm(\pm p)_{A_1\dots A_n}}{
[t_1^{b_1}\dots t_n^{b_n}p_{ b_1}\dots p_{ b_n}]
^{1/2}}
=
\frac{\pi_{ A_1}\dots \pi_{ A_n}f_\pm(\pm p)}{
[t_1^{b_1}\dots t_n^{b_n}p_{ b_1}\dots p_{ b_n}]
^{1/2}}.\label{wst}
\ee
The simplest and most straightforward way of getting 
 the Bargmann-Wigner amplitude directly from the spinor field 
is to choose  $t_k^a$ 
$p$-dependent and {\it null\/}, for example
\be
t_k^a=\omega^A \bar \omega^{A'},
\ee
where $\omega^A$ is the element of the spin-frame associated
with $p^a$. We extract the amplitude from the Fourier transform
by 
\be
f_\pm(\pm p)=
\omega^{A_1}\dots\omega^{A_n}
\psi_\pm(\pm p)_{A_1\dots A_n}.
\ee
It will be shown in the next section that the choice of {\it
null\/} and $p$-dependent $t^a$ is natural also in the massive case.

\section{Massive Bargmann-Wigner Fields}
\label{Mn0}

The Bargmann-Wigner equations \cite{BW,BR,Ohnuki} 
representing free spin-$n/2$
fields with mass $m\neq 0$ are equivalent to the set of spinor
field equations for $2^n$ fields $\psi_{A_1\dots A_n}^{0\dots
0}$,  $\psi_{A_1\dots A'_n}^{0\dots 1}$,... 
$\psi_{A'_1\dots A'_n}^{1\dots
1}$, 
\begin{eqnarray}
i\nabla{^A}{_{A'}}\psi(x)^{\dots 0\dots}_{\dots A\dots}
&=&-\frac{m}{\sqrt{2}} \psi(x)^{\dots 1\dots}_{\dots A'\dots},\\
i\nabla{_A}{^{A'}}\psi(x)^{\dots 1\dots}_{\dots A'\dots}
&=&\frac{m}{\sqrt{2}} \psi(x)^{\dots 0\dots}_{\dots A\dots}.
\end{eqnarray}
The convention we use differs slightly from the one introduced 
by Penrose and Rindler \cite{PR} (see Appendix~\ref{A2}).
The Fourier representation of the field is
\be
\psi(x)^{\dots}_{\dots}=
\frac{1}{(2\pi)^3}\int\frac{d^3p}{2p^0}\,
\Bigl\{
e^{-i p\cdot x}
\psi_+( p)^{\dots}_{\dots}
+
e^{i p\cdot x}
\psi_-(- p)^{\dots}_{\dots}
\Bigr\}
\ee
where $\psi_\pm( p)^{\dots}_{\dots}$ satisfy
\begin{eqnarray}
\pm p{^A}{_{A'}}\psi_\pm(\pm p)^{\dots 0\dots}_{\dots A\dots}
&=&
-\frac{m}{\sqrt{2}} \psi_\pm(\pm p)
^{\dots 1\dots}_{\dots A'\dots},\\
\pm p{_A}{^{A'}}\psi_\pm(\pm p)^{\dots 1\dots}_{\dots A'\dots}
&=&\frac{m}{\sqrt{2}} \psi_\pm(\pm p)^{\dots 0\dots}_{\dots A\dots}.
\end{eqnarray}
Consider now the tensor
\be
T_\pm( p)_{a_1\dots a_n}&=&
\psi_\pm( p)_{A_1\dots A_n}^{0\dots 0}
\bar \psi_\pm( p)_{A'_1\dots A'_n}^{0\dots 0}
+
\psi_\pm( p)_{A_1\dots A'_n}^{0\dots 1}
\bar \psi_\pm( p)_{A'_1\dots A_n}^{0\dots 1}
+\dots
+
\psi_\pm( p)_{A'_1\dots A'_n}^{1\dots 1}
\bar \psi_\pm( p)_{A_1\dots A_n}^{1\dots 1}\\
&=&
\psi_\pm( p)_{A_1\dots A_n}^{0\dots 0}
\overline{\psi_\pm( p)_{A_1\dots A_n}^{0\dots 0}}
+
\psi_\pm( p)_{A_1\dots A'_n}^{0\dots 1}
\overline{\psi_\pm( p)_{A_1\dots A'_n}^{0\dots 1}}
+\dots
+
\psi_\pm( p)_{A'_1\dots A'_n}^{1\dots 1}
\overline{\psi_\pm( p)_{A'_1\dots A'_n}^{1\dots 1}}.
\ee
The standard 
Bargmann-Wigner scalar product is defined by the norm
\be
\parallel\psi_\pm\parallel^2=
\int\frac{d^3p}{2|p^0|^{n+1}}\,\bigl\{
\psi_\pm(\pm p)_{0\dots 0}^{0\dots 0}
\overline{\psi_\pm(\pm p)_{0\dots 0}^{0\dots 0}}
+
\dots
+
\psi_\pm(\pm p)_{1'\dots 1'}^{1\dots 1}
\overline{\psi_\pm(\pm p)_{1'\dots 1'}^{1\dots 1}}\bigr\}
\ee
which being invariant under the Poincar\'e group
is not {\it manifestly\/} invariant. The lack of manifest 
invariance leads to difficulties with applying the 
spinor methods in the context of induced representations. 

To get the manifestly invariant form we shall first rewrite the
tensor $T_\pm( p)_{a_1\dots a_n}$ with the help of the
field equations as follows
\be
T_\pm(\pm p)_{a_1\dots a_k\dots a_n}=
2m^{-2}
p{_{A_k}}{^{B'_k}}p{^{B_k}}{_{A'_k}}
T_\pm(\pm p)_{a_1\dots b_k\dots a_n}=
2m^{-2}
\Bigl(
p_{ a_k}p^{b_k}-\frac{m^2}{2}g{_{a_k}}{^{b_k}}
\Bigr)
T_\pm(\pm p)_{a_1\dots b_k\dots a_n},
\ee
where we have used the trace-reversal identity.

Therefore
\be
T_\pm(\pm p)_{a_1\dots a_k\dots a_n}=
m^{-2}
p_{ a_k}p^{b_k}
T_\pm(\pm p)_{a_1\dots b_k\dots a_n}.\label{id}
\ee
Applying (\ref{id}) to itself $n$ times we get
\be
T_\pm(\pm p)_{a_1\dots a_n}=
m^{-2n}
p_{ a_1}\dots p_{ a_n}p^{b_1}\dots p^{b_n}
T_\pm(\pm p)_{b_1\dots b_n}.
\ee
The Poincar\'e (i.e. spinor) transformation of the
Bargmann-Wigner field implies
\be
T'_\pm(\pm p)_{a_1\dots a_n}&=&
m^{-2n}
p_{ a_1}\dots p_{ a_n}p^{b_1}\dots p^{b_n}
T'_\pm(\pm p)_{b_1\dots b_n}\noindent\\
&=&
m^{-2n}
p_{ a_1}\dots p_{ a_n}p^{b_1}\dots p^{b_n}
\Lambda{_{b_1}}{^{c_1}}\dots\Lambda{_{b_n}}{^{c_n}}
T_\pm(\pm {\Lambda^{-1}p})_{c_1\dots c_n}\noindent\\
&=&
m^{-2n}
p_{ a_1}\dots p_{ a_n}
(\Lambda^{-1}p)^{b_1}
\dots(\Lambda^{-1}p)^{b_n}
T_\pm(\pm {\Lambda^{-1}p})_{b_1\dots b_n}
\ee
Let $t_1^{a},\dots,t_n^{a}$ be arbitrary 
world-vectors satisfying $t_k^{a}p_{a}\neq 0$ for any
future-pointing $p_a$ 
belonging to the mass hyperboloid
(the vectors can be $p$-dependent).
The expression 
\be
\parallel \psi_\pm\parallel'^2=
\int\frac{d^3p}{2p^0}
\frac{
t_1^{a_1}\dots t_n^{a_n}
T_\pm(\pm p)_{a_1\dots a_n}}{
t_1^{b_1}\dots t_n^{b_n}p_{ b_1}\dots p_{ b_n}}=
m^{-2n}\int\frac{d^3p}{2p^0}
p^{a_1}\dots p^{a_n}
T_\pm(\pm p)_{a_1\dots a_n}\label{ex}
\ee
is manifestly invariant. The LHS of
(\ref{ex}) is independent of the choice of
$t_1^{a},\dots,t_n^{a}$ because the RHS is independent of them.

We can write also an analog of (\ref{gexx}) for the massive
fields:
\be
\parallel \psi_\pm\parallel'^2=
\int d\mu_{t,m}(^tp^a)
\frac{
t_1^{a_1}\dots t_n^{a_n}
{^t}T_\pm(\pm ^tp^a)_{a_1\dots a_n}}{
t_1^{b_1}\dots t_n^{b_n}p_{ b_1}\dots p_{ b_n}}=
m^{-2n}\int d\mu_{t,m}(^tp^a)
p^{a_1}\dots p^{a_n}
{^t}T_\pm(\pm ^tp^a)_{a_1\dots a_n}.
\ee
Analogously to the massless case we find 
\be
\parallel \psi_\pm\parallel'^2=
\int\frac{d^3p}{2|p^0|^{n+1}}
T_\pm(\pm p)_{0\dots 0}=
2^{-n/2}
\parallel \psi_\pm\parallel^2.
\ee

The massive fields 
$\psi_{A_1\dots A_n}^{0\dots
0}$,  $\psi_{A_1\dots A'_n}^{0\dots 1}$,... 
$\psi_{A'_1\dots A'_n}^{1\dots
1}$ as they stand are not necessarily symmetric in all indices,
so correspond to representations which are in general reducible.
To get irreducible fields we assume their symmetry in
all indices. The number of the fields is $n+1=2j+1$, where
$j$ is the spin of the Bargmann-Wigner field. Choose now 
the $p$-dependent $t^a=\omega^a$ used in Appendix~\ref{APL}. 
Using the normalization (\ref{normalization}) we get from (\ref{ex})
\be
\parallel \psi_\pm\parallel'^2&=&
\int\frac{d^3p}{2p^0}
\omega^{a_1}\dots \omega^{a_n}
T_\pm(\pm p)_{a_1\dots a_n}
(\omega\cdot p)^{-n}\nonumber\\&=&
\int\frac{d^3p}{2p^0}
\Bigl(
|f_\pm(\pm p)^{0\dots 0}|^2+|f_\pm(\pm p)^{0\dots 01}|^2+
\dots + |f_\pm(\pm p)^{1\dots 1}|^2
\Bigr)
\label{exBW}
\ee
where (see Appendix~\ref{Awave})
\be
N(\omega,\pi)^{n}
f_\pm(\pm p)\overbrace{^{0\dots 0}}^r\overbrace{^{1\dots
1}}^{n-r}= 
\omega^{A_1}\dots \omega^{A_r}
\bar \omega^{A'_{r+1}}\dots \bar \omega^{A'_n}
\psi_\pm(\pm p)^{0\dots 01\dots
1}_{A_1\dots A_rA'_{r+1}\dots A'_n}.\label{wave}
\ee

\section{Acknowledgements}

The results presented in this series of papers
have their roots in many hours of extensive
discussions with Gerry Kaiser on position operators and 
phase-space reformulation of relativistic quantum theory.
I am indebted to  Gerry Kaiser and Dave
Pritchard for their hospitality and support at University of
Massachusetts-Lowell and Massachusetts Institute of Technology
where a part of this work was done. 
My work at UML and MIT was made possible
by the Fulbright Commission and is a part of
the KBN project 2P30B03809.

\section{Appendices}

\subsection{Infeld-van der Waerden tensors and generators of
(1/2,0) and (0,1/2)} \label{A1}

Consider  representations $(\frac{1}{2},0)$ and 
 $(0,\frac{1}{2})$
 of an element $\omega\in SL(2,C)$: $e^{\frac{i}{2}
\omega^{ab}{\sigma}_{ab}}$ and 
$e^{\frac{i}{2}
 \omega^{ab}\bar {\sigma}_{ab}}$.
 The explicit form of the generators in terms of 
Infeld-van der Waerden tensors is
\begin{eqnarray}
\frac{1}{2i}\Bigl(g^a_{\pp A XA'}g^{bYA'}-g^b_{\pp A
XA'}g^{aYA'}\Bigr)
&=&{ \sigma}^{ab\pp A
Y}_{\pp {aa}X},\\
\frac{1}{2i}\Bigl(
g^a_{\pp A AX'}g^{bAY'}-g^b_{\pp A AX'}g^{aAY'}\Bigr)&=&\bar
{ \sigma}^{ab\pp A
Y'}_{\pp {aa}X'}.
\end{eqnarray}
Their purely spinor form is
\begin{eqnarray}
{\sigma}_{AA'BB'XY}&=&\frac{1}{2i}\varepsilon
_{A'B'}\bigl(\varepsilon_{AX}\varepsilon_{BY}+
\varepsilon_{BX}\varepsilon_{AY}\bigr),\\
\bar {\sigma}_{AA'BB'X'Y'}
&=&\frac{1}{2i}\varepsilon_{AB}\bigl(\varepsilon_{A'X'}\varepsilon_{B'Y'}+
\varepsilon_{B'X'}\varepsilon_{A'Y'}\bigr),
\end{eqnarray}
Dual tensors are $^*\bar
{\sigma}^{ab\pp A
Y'}_{\pp {aa}X'}=+i\bar
{\sigma}^{ab\pp A
Y'}_{\pp {aa}X'}$
and 
$^*{\sigma}^{ab\pp A
Y}_{\pp {aa}X}=-i{\sigma}^{ab\pp A
Y}_{\pp {aa}X}$.

Additionally the Infeld-van der Waerden tensors satisfy
\begin{eqnarray}
g^a_{\pp A XA'}g^{bYA'}+g^b_{\pp A
XA'}g^{aYA'}
&=&g^{ab}\varepsilon^{{\pp {X}}
Y}_{X}\label{IW1}\\
g^a_{\pp A AX'}g^{bAY'}+g^b_{\pp A
AX'}g^{aAY'}
&=&g^{ab}\varepsilon^{{\pp {X}}
Y'}_{X'}\label{IW2}
\end{eqnarray}
These equations lead to the identities
\be
g^a_{\pp A XA'}g^{bYA'}&=&\frac{1}{2}g^{ab}\varepsilon^{{\pp {X}}
Y}_{X}+i{ \sigma}^{ab\pp A
Y}_{\pp {aa}X}\label{id1}\\
g^a_{\pp A AX'}g^{bAY'}&=&\frac{1}{2}g^{ab}\varepsilon^{{\pp {X}}
Y'}_{X'}+i
\bar
{ \sigma}^{ab\pp A
Y'}_{\pp {aa}X'}\label{id2}
\ee

\subsection{Gamma matrices and the Dirac equation in the spinor form}
\label{A2}

The 2-spinor form of the Dirac equation given in \cite{PR} leads
to difficulties with a direct comparison of spinor formulas
with standard quantum mechanics textbooks. This appendix
explains the convention used in this paper.

The Dirac equation in the momentum representation 
can be written in a matrix form as follows
\be
\pm
\left(
\begin{array}{cc}
0 & (p^0 + \bbox p\cdot\bbox \sigma){^{\bf A}}{^{{\bf B}'}}\\
 (p^0 - \bbox p\cdot\bbox \sigma){_{{\bf A}'}}{_{\bf B}} & 0
\end{array}
\right)
\left(
\begin{array}{c}
\psi_\pm(\pm p)^{\bf B}\\
\xi_\pm(\pm)_{{\bf B}'}
\end{array}
\right)
&=&
\left(
\begin{array}{cc}
0 & \pm p^a\sigma_a{^{\bf A}}{^{{\bf B}'}}\\
\pm p^a\tilde \sigma_a{_{{\bf A}'}}{_{\bf B}} & 0
\end{array}
\right)
\left(
\begin{array}{c}
\psi_\pm(\pm p)^{\bf B}\\
\xi_\pm(\pm p)_{{\bf B}'}
\end{array}
\right)\nonumber\\
&=&m
\left(
\begin{array}{c}
\psi_\pm(\pm p)^{\bf A}\\
\xi_\pm(\pm p)_{{\bf A}'}
\end{array}
\right),
\ee
where
\be
\sigma_a{^{\bf A}}{^{{\bf B}'}}&=&( 1, \bbox \sigma){^{\bf A}}
{^{{\bf B}'}}\\
\tilde \sigma_a{_{{\bf A}'}}{_{\bf B}}
&=&( 1,- \bbox \sigma){_{{\bf A}'}}{_{\bf B}},
\ee
and $\bbox\sigma$ is a matrix vector whose components are the
Pauli matrices. The following
identification with the Infeld-van der Waerden symbols can be
made
\be
g_{\bf a}{_{\bf A}}{_{{\bf B}'}}&=&\frac{1}{\sqrt{2}}\tilde
\sigma_{\bf a}{_{{\bf B}'}}{_{\bf A}}\\ 
g_{\bf a}{^{\bf A}}{^{{\bf B}'}}
&=&\frac{1}{\sqrt{2}}
\sigma_{\bf a}{^{\bf A}}{^{{\bf B}'}}
\ee
The Dirac equation in the Minkowski space
representation is ($\hbar=1$)
\begin{eqnarray}
i\nabla_{AA'}\psi^A&=&\frac{m}{\sqrt{2}} \xi_{A'},\\
i\nabla^{AA'}\xi_{A'}&=&\frac{m}{\sqrt{2}} \psi^{A}
\end{eqnarray}
where $\nabla_{AA'}=\nabla^ag_{aAA'}$ etc. This equation differs
by a sign and the presence of $i$ from the 
form given in \cite{PR}. 
The matrix form of the equation 
\be
\left(
\begin{array}{cc}
0 & \pm p{_{A}}{^{B'}}\\
\mp p{^{B}}{_{A'}} & 0
\end{array}
\right)
\left(
\begin{array}{c}
\psi_\pm(\pm p)_B\\
\xi_\pm(\pm p)_{B'}
\end{array}
\right)
=\frac{m}{\sqrt{2}}
\left(
\begin{array}{c}
\psi_\pm(\pm p)_A\\
\xi_\pm(\pm p)_{A'}
\end{array}
\right)
\ee
shows that the Dirac gamma matrices are given by
\be
\gamma{_q}{_\alpha}{^\beta}=
\sqrt{2}\left(
\begin{array}{cc}
0 & g{_{qA}}{^{B'}}\\
-g{_q}{^B}{_{A'}} & 0
\end{array}
\right)
\ee
Product of two gamma matrices
\be
\gamma{_q}{_\alpha}{^\beta}\gamma{_r}{_\beta}{^\gamma}&=&
\left(
\begin{array}{cc}
g{_{qr}}\varepsilon{_A}{^C}+2i\sigma{_{qr}} {_A}{^C} & 0\\
0 & g{_{qr}}\varepsilon{_{A'}}{^{C'}}+2i\bar \sigma{_{qr}}
{_{A'}}{^{C'}}
\end{array}
\right)=
g{_{qr}}I{_\alpha}{^\gamma}+2i\sigma{_{qr}} {_\alpha}{^\gamma}
\ee
implies
\be
\gamma{_q}{_\alpha}{^\beta}\gamma{_r}{_\beta}{^\gamma}+
\gamma{_r}{_\alpha}{^\beta}\gamma{_q}{_\beta}{^\gamma}
&=&2g{_{qr}}I{_\alpha}{^\gamma}\label{znak1},\\
\gamma{_q}{_\alpha}{^\beta}\gamma{_r}{_\beta}{^\gamma}-
\gamma{_r}{_\alpha}{^\beta}\gamma{_q}{_\beta}{^\gamma}
&=&4i\sigma{_{qr}} {_\alpha}{^\gamma}.\label{znak2}
\ee
(\ref{znak2}) differs by the factor $(-1/2)$ 
from the definition from \cite{BD}
 because there the  generators are defined 
by $S(\omega)=e^{-\frac{i}{4}
\omega^{ab}{\sigma}_{ab}}$. There is also a difference with
respect to \cite{PR} where the gamma matrices
are defined 
 without the $-$
sign  (this would
lead to the opposite sign at the RHS of (\ref{znak1})). 

The spinor form of the Dirac current is 
\be
j_a=\sqrt{2}g_a{^{AA'}}\bigl(\psi_A\bar \psi_{A'} +
\xi_{A'}\bar \xi_{A}\bigr).\label{spin-cur}
\ee
(\ref{spin-cur}) is derived bispinorially as follows
\be
j_a=
\sqrt{2}(\bar \psi^{A'},\bar \xi^A)
\left(
\begin{array}{cc}
0 & \varepsilon{_{A'}}{^{B'}}\\
-\varepsilon{_A}{^{B}} & 0
\end{array}
\right)
\left(
\begin{array}{cc}
0 & g{_{aB}}{^{C'}}\\
-g{_a}{^C}{_{B'}} & 0
\end{array}
\right)
\left(
\begin{array}{c}
\psi_C\\
\xi_{C'}
\end{array}
\right)
\ee
and
\be
j_0&=&
(\bar \psi^{A'},\bar \xi^A)
\underbrace{
\left(
\begin{array}{cc}
0 & \varepsilon{_{A'}}{^{B'}}\\
-\varepsilon{_A}{^{B}} & 0
\end{array}
\right)}_{``\gamma_0"}
\underbrace{\sqrt{2}
\left(
\begin{array}{cc}
0 & g{_{0B}}{^{C'}}\\
-g{_0}{^C}{_{B'}} & 0
\end{array}
\right)}_{``\gamma_0"}
\left(
\begin{array}{c}
\psi_C\\
\xi_{C'}
\end{array}\right)
\ee
showing that the matrix $\gamma_0$ appearing in textbooks
corresponds actually to two different spinor objects.
The pseudoscalar matrix $\gamma_5$ corresponds to the spinor
matrix 
\be
\gamma{_5}{_X}{^Y}&=&
\frac{i}{4!}
e^{abcd}\gamma_a\gamma_b\gamma_c\gamma_d{_X}{^Y}
=
\left(
\begin{array}{cc}
-\ve{_X}{^Y} & 0\\
0 &  \ve{_{X'}}{^{Y'}}
\end{array}
\right).
\ee

\subsection{Spinor form of the Pauli-Lubanski vector}
\label{APL}

The author of this paper is not aware of any work where
an explicitly covariant presentation of the spinor form of the
P-L vector and its eigenvalue problem could be found. This
appendix contains basic formulas and explains in detail the
``null formalism" used in expanding solutions of the
Bargmann-Wigner equations in terms of eigenvectors of the P-L
vector's projections in null directions. 

Let $P^a$ denote generators of the Minkowski space 4-translations.
The P-L vectors corresponding to $(1/2,0)$ and $(0,1/2)$
representations are defined by
\be
S^a{_{X}}{^{Y}}&=&P_b{^*}{\sigma}^{ba\pp A Y}_{\pp {aa}X},\\
S^a{_{X'}}{^{Y'}}&=&P_b{^*}{\bar \sigma}^{ba\pp A Y'}_{\pp
{aa}X'}.
\ee
Their momentum representation is 
\begin{eqnarray}
S^a(p){_{X}}{^{Y}}&=&
-\frac{1}{2}\Bigl(p{_{XX'}}g^{aYX'}-g^a_{\pp A
XX'}p^{YX'}\Bigr)=
-\frac{1}{2}\Bigl(p{_{X}}{^{A'}}\varepsilon^{AY}
-\varepsilon{_{X}}{^A}p^{YA'}\Bigr)
\\
S^a(p){_{X'}}{^{Y'}}&=&
\frac{1}{2}\Bigl(p{_{ AX'}}g^{aAY'}-g^a_{\pp A
AX'}p^{AY'}\Bigr)
=
\frac{1}{2}\Bigl(p{^{A}}{_{X'}}\varepsilon^{A'Y'}
-\varepsilon{_{X'}}{^{A'}}p^{AY'}\Bigr)
\end{eqnarray}
Using the trace-reversal formula, the identity
\be
p_{AA'}p^{AB'}=\frac{1}{2}p_a p^a\, \varepsilon{_{A'}}{^{B'}},
\ee
and its complex-conjugated version, we get in the massless case
\be
S^a( p){_{X}}{^{Y}}p{_{ YX'}}&=&
-\frac{1}{2}p^a \,p{_{ XX'}},\label{pl0}\\
S^a( p){_{X'}}{^{Y'}}p{_{ XY'}}&=&
\frac{1}{2}p^a \,p{_{ XX'}},\label{pl0'}
\ee
which imply (\ref{r,s}), which  means
that the spinor
$
\xi_\pm( p)^{A'_1\dots A'_n}
$
in (\ref{pH}) is in fact arbitrary. 

The massive case is more complicated since the components of the
P-L vector no longer commute. Consider a projection
of the P-L
 vector in the direction of a (timelike, spacelike or
null, and generally $p$-dependent) world-vector $t^a$ 
\be
S(t,p){_{X}}{^{Y}}=t_aS^a( p){_{X}}{^{Y}}=t\cdot S( p){_{X}}{^{Y}}=
\frac{1}{2}\Bigl(t{^{Y}}{_{X'}}p{_{X}}{^{X'}}
+t{_{XX'}}p^{YX'}\Bigr),\\
S(t,p){_{X'}}{^{Y'}}=
t_aS^a( p){_{X'}}{^{Y'}}=t\cdot S( p){_{X'}}{^{Y'}}=
-\frac{1}{2}\Bigl(t{_{X}}{^{Y'}}p{^{X}}{_{X'}}
+t{_{XX'}}p^{XY'}\Bigr).
\ee
To find eigenvalues of these operators let us first observe  that
if $S_{XY}$ is symmetric then 
\be
S{_X}{^Z}S{_Z}{^Y}=-\frac{1}{2}S_{AB}S^{AB}\varepsilon{_X}{^Y}
\ee
and hence the eigenvalues of $S{_X}{^Y}$ are $\pm \bigl[
-\frac{1}{2}S_{AB}S^{AB}\bigr]^{1/2}$. 
Projectors projecting on the eigenstates are
\be
\Pi^{(\pm)}{_A}{^B}=
\frac{1}{2}\Bigl(
\varepsilon{_A}{^B}\pm \Bigl[-\frac{1}{2}S_{XY}S^{XY}\Bigr]^{-1/2}
S{_A}{^B}\Bigr).
\ee
Analogous formulas hold for symmetric spinors with two primed
indices. 

Applying these results to the projection of the
P-L vector in the direction $t^a$  we find eigenvalues
\be
\frac{1}{2}\lambda^{(\pm)}(t,p)&=&\pm \Bigl[
-\frac{1}{2}S(t,p)_{AB}S(t,p)^{AB}\Bigr]^{1/2}=
\pm \Bigl[
-\frac{1}{2}S(t,p)_{A'B'}S(t,p)^{A'B'}\Bigr]^{1/2}\nonumber\\
&=&\pm \frac{1}{2}\sqrt{(t\cdot p)^2 -m^2t^2 }.\label{plev}
\ee
Formula (\ref{plev}) shows that there exists a privileged choice
of $t^a$, namely future-pointing and {\it null\/}. Indeed, for
future pointing $t^a,\,p^a$ we get 
\be
 \frac{1}{2}\lambda^{(\pm)}(t,p)
&=&\pm \frac{1}{2}t\cdot p \label{plev'}
\ee
which is analogous to the massless case even though, in general,
$m^2\neq 0$ in (\ref{plev}). The corresponding eigenstates are
determined by the projectors
\be
\Pi^{(\pm)}(t,p){_A}{^B}&=&
\frac{1}{2}\Bigl(
\varepsilon{_A}{^B} + \frac{2}{\lambda^{(\pm)}}
S(t,p){_A}{^B}\Bigr),\\
\Pi^{(\pm)}(t,p){_{A'}}{^{B'}}&=&
\frac{1}{2}\Bigl(
\varepsilon{_{A'}}{^{B'}} + \frac{2}{\lambda^{(\pm)}}
S(t,p){_{A'}}{^{B'}}\Bigr).
\ee
The projectors
\be
P(\pm p){_\alpha}{^\beta}=
\frac{1}{2}
\left(
\begin{array}{cc}
\varepsilon{_A}{^B} & \pm\frac{\sqrt{2}}{m} p{_{A}}{^{B'}}\\
\mp\frac{\sqrt{2}}{m} p{^{B}}{_{A'}} & \varepsilon{_{A'}}{^{B'}}
\end{array}
\right)
\ee
project solutions of the Dirac equation on positive ($+$) and
negative ($-$) energy states i.e.
\be
P(\pm p){_\alpha}{^\beta}\Psi_{\pm\beta}(\pm p)=
\Psi_{\pm\alpha}(\pm p).
\ee
The useful formula
\be
S^a{_A}{^B}p{_B}{^{B'}}=
p{_A}{^{A'}}S^a{_{A'}}{^{B'}}.\label{useful}
\ee
implies that 
\be
P(\pm p){_\alpha}{^\beta}S^a(p){_\beta}{^\gamma}=
S^a(p){_\alpha}{^\beta}P(\pm p){_\beta}{^\gamma},
\ee
where
\be
S^a(p){_\alpha}{^\beta}=
\left(
\begin{array}{cc}
S^a(p){_A}{^B} & 0\\
0 & S^a(p){_{A'}}{^{B'}}
\end{array}
\right).
\ee
Let
\be
\Pi^{(\pm)}(t,p){_{\alpha}}{^{\beta}}=
\left(
\begin{array}{cc}
\Pi^{(\pm)}(t,p){_{A}}{^{B}} & 0\\
0 & \Pi^{(\pm)}(t,p){_{A'}}{^{B'}}
\end{array}
\right).
\ee
Then
\be
\Pi^{(\pm)}(t,p){_{\alpha}}{^{\beta}}
P(\pm p){_\beta}{^\gamma}&=&
P(\pm p){_\alpha}{^\beta}
\Pi^{(\pm)}(t,p){_{\beta}}{^{\gamma}}=
\Pi^{(\pm)}_{\pm}(t,p){_{\alpha}}{^{\gamma}}\label{PPi}\\
&=&
\frac{1}{4\lambda^{(\pm)}}
\left(
\begin{array}{rl}
\lambda^{(\pm)}\varepsilon{_A}{^C}
+ t{^C}{_{X'}}p{_{A}}{^{X'}} + t{_{AX'}}p{^{CX'}}, & 
\pm\frac{\sqrt{2}}{m}\bigl[(\lambda^{(\pm)}-t\cdot p) 
p{_{A}}{^{C'}}+ m^2 t{_{A}}{^{C'}}\bigr]\\
\mp\frac{\sqrt{2}}{m}\bigl[(\lambda^{(\pm)}+t\cdot p) 
p{^{C}}{_{A'}}- m^2 t{^{C}}{_{A'}}\bigr], & 
\lambda^{(\pm)}\varepsilon{_{A'}}{^{C'}}
- t{_X}{^{C'}}p{^{X}}{_{A'}} - t{_{XA'}}p{^{XC'}}
\end{array}
\right)\nonumber
\ee
where $\lambda^{(\pm)}=\lambda^{(\pm)}(t,p)$. The signs ``$\pm$"
of energy are independent of the signs ``$(\pm)$" of spin. 

To simplify the form of (\ref{PPi}) 
consider a future-pointing and non-null energy-momentum vector
$p^a$, and a future-pointing and
 null vector $\omega^a$ satisfying $p\cdot
\omega=m/\sqrt{2}$ (such an $\omega^a$ always exists). 
The difference 
\be
\frac{m}{\sqrt{2}}\pi^a=p^a-\frac{m}{\sqrt{2}}\omega^a
\ee
is also  null and future-pointing,  $\omega\cdot\pi=1$, and
it follows that
\be
p^a=
\frac{m}{\sqrt{2}}\Bigl(\omega^{a}+
\pi^{a}
\bigr)
=
\frac{m}{\sqrt{2}}\Bigl(\omega^{A} \bar \omega^{A'}+
\pi^{A} \bar \pi^{A'}
\bigr).\label{p}
\ee
The spinors $\pi^{A}$ and $\omega^{A}$ can be explicitly
constructed as follows. Choose an arbitrary $p$-independent
spinor $\nu^A$. Let $n^a=\nu^A \bar \nu^{A'}$. We know that 
$n\cdot p$ is never vanishing if $p^2=m^2> 0$. Define
\be
\omega^{A}
&=&\Bigl[\frac{m}{\sqrt{2}}\Bigr]^{1/2}\frac{\nu^A}{
\sqrt{p^{BB'}\nu_B
\bar \nu_{B'}}}=\omega^{A}(\nu,p)\\
\pi^A 
&=&\Bigl[\frac{\sqrt{2}}{m}\Bigr]^{1/2}
\frac{p^{AA'}\bar \nu_{A'}}
{\sqrt{p^{BB'}\nu_B
\bar \nu_{B'}}}=\pi^A(\nu, p),
\ee
which satisfy
$
\omega_A\pi^A=1,
$
$\omega
\cdot p=m/\sqrt{2}$
and 
\be
\pi^a&=&\pi^A\bar \pi^{A'}=\frac{\sqrt{2}}{m}p^{AB'}p^{BA'}
\nu_B\bar \nu_{B'}\frac{1}{p^{CC'}\nu_C
\bar \nu_{C'}}=
\frac{\sqrt{2}}{m}\Bigl(p^{a}p^{b}-\frac{m^2}{2}g^{ab}\Bigr)
n_b\frac{1}{p^{CC'}\nu_C
\bar \nu_{C'}}\nonumber\\
&=&
\frac{\sqrt{2}}{m}p^a - \omega^a
\ee
as required.
Let us take $t^a=\omega^a$. The eigenvalues of the P-L vector in
the $t^a$ direction are now
\be
\lambda^{(\pm)}(t,p)=\pm t\cdot p=\pm \frac{m}{\sqrt{2}}.
\ee
The projectors $\Pi^{(\pm)}_{\pm}(t,p)$ are
\be
\Pi^{(+)}_{\pm}(t,p){_{\alpha}}{^{\gamma}}&=&
\frac{1}{4}
\left(
\begin{array}{cc}
\varepsilon{_A}{^C}
+\pi{_{A}}\omega{^C} + \omega{_A}\pi{^{C}} & 
\pm\, 2\,\omega_{A}\bar \omega^{C'}  \\
\mp\, 2\,\bar \pi_{A'} \pi^{C}&
\varepsilon{_{A'}}{^{C'}}
-\bar \pi{_{A'}}\bar \omega{^{C'}} - \bar \omega{_{A'}}
\bar \pi{^{C'}}
\end{array}
\right),\\
\Pi^{(-)}_{\pm}(t,p){_{\alpha}}{^{\gamma}}&=&
\frac{1}{4}
\left(
\begin{array}{cc}
\varepsilon{_A}{^C}
-\pi{_{A}}\omega{^C} - \omega{_A}\pi{^{C}} & 
\pm\, 2\,\pi_{A}\bar \pi^{C'}  \\
\mp\, 2\,\bar \omega_{A'} \omega^{C}&
\varepsilon{_{A'}}{^{C'}}
+\bar \pi{_{A'}}\bar \omega{^{C'}} + \bar \omega{_{A'}}
\bar \pi{^{C'}}
\end{array}
\right).
\ee
Return for a moment to the massless case. Let $p^a=\pi^A\bar
\pi^{A'}$, and let $\omega^A$ be a spin-frame partner of 
$\pi^A$, i.e. $\pi_A\omega^A=1$. Let $t^a=\omega^A\bar
\omega^{A'}$. Then 
\be
S(t,p){_{A}}{^{B}}&=&
-\frac{1}{2}\Bigl(
\pi_{A}\omega^{B}+ \omega_{A}\pi^{B}\Bigr),\\
S(t,p){_{A'}}{^{B'}}&=&
\frac{1}{2}\Bigl(
\bar \pi_{A'}\bar \omega^{B'}+ 
\bar \omega_{A'}\bar \pi^{B'}\Bigr).
\ee
The eigenequations (\ref{pl0}), (\ref{pl0'}) are equivalent to 
\be
S(t,p){_{A}}{^{B}}\pi_{B}&=&
-\frac{1}{2}\pi_{A},\label{pl00}\\
S(t,p){_{A'}}{^{B'}}\bar \pi_{B'}&=&
\frac{1}{2}\bar \pi_{A'},\label{pl00'}
\ee
which can be supplemented by 
\be
S(t,p){_{A}}{^{B}}\omega_{B}&=&
\frac{1}{2}\omega_{A},\label{pl000}\\
S(t,p){_{A'}}{^{B'}}\bar \omega_{B'}&=&
-\frac{1}{2}\bar \omega_{A'}.\label{pl000'}
\ee
In the massive case, for $t^a=\omega^a$, we find analogous
formulas 
\be
S(t,p){_{A}}{^{B}}&=&
\frac{1}{2}t\cdot p\Bigl(
\omega_{A}\pi^{B}+ \pi_{A}\omega^{B}\Bigr),\\
S(t,p){_{A'}}{^{B'}}&=&
-\frac{1}{2}t\cdot p\Bigl(
\bar \omega_{A'}\bar \pi^{B'}+ 
\bar \pi_{A'}\bar \omega^{B'}\Bigr),
\ee
and
\be
S(t,p){_{A}}{^{B}}\omega_{B}&=&
\frac{1}{2}t\cdot p\,\omega_{A},\\
S(t,p){_{A'}}{^{B'}}\bar \omega_{B'}&=&
-\frac{1}{2}t\cdot p\,\bar \omega_{A'},\\
S(t,p){_{A}}{^{B}}\pi_{B}&=&
-\frac{1}{2}t\cdot p\,\pi_{A},\\
S(t,p){_{A'}}{^{B'}}\bar \pi_{B'}&=&
\frac{1}{2}t\cdot p\,\bar \pi_{A'}.
\ee
Let 
\be
\Pi^{(\pm)}_{\pm}(t,p){_\alpha}{^\beta}\chi{^{(\pm)}_{\pm}}_
\beta=\chi{^{(\pm)}_{\pm}}_\alpha.
\ee
We can take
\be
\chi{^{(+)}_{\pm\alpha}}&=&
N(\omega,\pi)
\left(
\begin{array}{c}
\pm\omega_{A}\\
-\bar \pi_{A'}
\end{array}
\right),\label{chi+}\\
\chi{^{(-)}_{\pm\alpha}}&=&
N(\omega,\pi)\left(
\begin{array}{c}
-\pi_{A}\\
\mp\bar \omega_{A'}
\end{array}
\right),\label{chi-}
\ee
where the scalar function
$N(\omega,\pi)$ depends on the choice of normalization.
The Dirac bispinor written in terms of
$\chi{^{(\pm)}_{\pm\alpha}}$ is
\be
\Psi_\pm(\pm p)_\alpha=
\left(
\begin{array}{c}
\psi_\pm(\pm p)^0_A\\
\psi_\pm(\pm p)^1_{A'}
\end{array}
\right)=
-N(\omega,\pi)\left(
\begin{array}{c}
\mp\omega_A\,f^{(+)}_\pm(\pm p)+\pi_A\,f^{(-)}_\pm(\pm p)\\
\bar \pi_{A'}\,f^{(+)}_\pm(\pm p)\pm\bar 
\omega_{A'}\,f^{(-)}_\pm(\pm p)
\end{array}
\right)
\ee
which implies 
\be
\bar \omega^{A'}\psi_\pm(\pm p)^1_{A'}&=&
N(\omega,\pi)f^{(+)}_\pm(\pm p)=N(\omega,\pi)f^{1}_\pm(\pm p),\\
\omega^{A}\psi_\pm(\pm p)^0_{A}&=&
N(\omega,\pi)f^{(-)}_\pm(\pm p)=N(\omega,\pi)f^{0}_\pm(\pm p).
\ee
In these formulas we have
redefined the sign-of-spin indices as follows: $(+)\to
1$, $(-)\to 0$. This convention is especially useful in
transition to Bargmann-Wigner fields of arbitrary spin. 
Taking $t^a=\omega^a$ in (\ref{ex}) we get for the Dirac equation
\be
\parallel \Psi_\pm\parallel'^2=
\int\frac{d^3p}{2p^0}
\frac{
\omega^{a}
T_\pm(\pm p)_{a}}{
\omega\cdot p}=
\int\frac{d^3p}{2p^0}
\frac{
|N(\omega,\pi)|^2
}{\omega\cdot p}\Bigl(|f^{0}_\pm(\pm p)|^2+|f^{1}_\pm(\pm
p)|^2 \Bigr).
\label{exD}
\ee
The simplest choice of normalization is 
\be
N(\omega,\pi)=\bigl[\omega\cdot p\bigr]^{1/2}=
\Bigl[\frac{m}{\sqrt{2}}\Bigr]^{1/2}\label{normalization}
\ee
implying
\be
\parallel \Psi_\pm\parallel'^2=
\int\frac{d^3p}{2p^0}
\Bigl(|f^{0}_\pm(\pm p)|^2+|f^{1}_\pm(\pm
p)|^2 \Bigr).
\ee
All the conventions used in this Appendix refer to
eigenstates of the P-L vector $S(t,+p)$ and all signs of spin would
be reversed if we had used $S(t,-p)$. 

\subsection{Proof of Eq.~(56)}
\label{Awave}

A Fourier transform of the spin-$n/2$ 
massive Bargmann-Wigner field can be written as a
linear combination of tensor products of the basic bispinors
(\ref{chi+}), (\ref{chi-}):
\be
\chi{^{(+)}_{\pm\alpha_1}}\dots \chi{^{(+)}_{\pm\alpha_n}}
&=&
N(\omega,\pi)^n
\left(
\begin{array}{c}
\chi{^{(+)}_{\pm(A_1}}\dots \chi{^{(+)}_{\pm A_n)}}\\
\chi{^{(+)}_{\pm(A_1}}\dots \chi{^{(+)}_{\pm A'_n)}}\\
\vdots\\
\chi{^{(+)}_{\pm(A'_1}}\dots \chi{^{(+)}_{\pm A'_n)}}
\end{array}
\right)=
N(\omega,\pi)^n
\left(
\begin{array}{c}
(\pm 1)^n\omega_{A_1}\dots \omega_{A_n}\\
(-1)(\pm 1)^{n-1}\omega_{A_1}\dots \omega_{A_{n-1}}\bar
\pi_{A'_n}\\ 
\vdots\\
(- 1)^n\bar \pi_{A'_1}\dots \bar \pi_{A'_n}
\end{array}
\right)\nonumber\\
\chi{^{(+)}_{\pm(\alpha_1}}\dots \chi{^{(-)}_{\pm\alpha_n)}}
&=&
N(\omega,\pi)^n
\left(
\begin{array}{c}
\chi{^{(+)}_{\pm(A_1}}\dots \chi{^{(-)}_{\pm A_n)}}\\
\chi{^{(+)}_{\pm(A_1}}\dots \chi{^{(-)}_{\pm A'_n)}}\\
\vdots\\
\chi{^{(+)}_{\pm(A'_1}}\dots \chi{^{(-)}_{\pm A'_n)}}
\end{array}
\right)=
N(\omega,\pi)^n
\left(
\begin{array}{c}
(\pm 1)^{n-1}(-1)\omega_{(A_1}\dots \omega_{A_{n-1}}\pi_{A_n)}\\
(\pm 1)^{n-1}(\mp 1)\omega_{A_1}\dots \omega_{A_{n-1}}\bar
\omega_{A'_n}\\ 
\vdots\\
(- 1)^n\bar \pi_{A'_1}\dots \bar \pi_{A'_{n-1}}\pi_{A_n}\\
(- 1)^{n-1}(\mp 1)\bar \pi_{(A'_1}\dots \bar \pi_{A'_{n-1}}
\bar \omega_{A'_n)}
\end{array}
\right)\nonumber\\
\vdots\nonumber\\
\chi{^{(-)}_{\pm\alpha_1}}\dots \chi{^{(-)}_{\pm\alpha_n}}
&=&
N(\omega,\pi)^n
\left(
\begin{array}{c}
\chi{^{(-)}_{\pm(A_1}}\dots \chi{^{(-)}_{\pm A_n)}}\\
\chi{^{(-)}_{\pm(A_1}}\dots \chi{^{(-)}_{\pm A'_n)}}\\
\vdots\\
\chi{^{(-)}_{\pm(A'_1}}\dots \chi{^{(-)}_{\pm A'_n)}}
\end{array}
\right)=
N(\omega,\pi)^n
\left(
\begin{array}{c}
(-1)^n\pi_{A_1}\dots \pi_{A_n}\\
(\mp 1)(- 1)^{n-1}\pi_{A_1}\dots \pi_{A_{n-1}}\bar
\omega_{A'_n}\\ 
\vdots\\
(\mp 1)^n\bar \omega_{A'_1}\dots \bar \omega_{A'_n}
\end{array}
\right).\nonumber
\ee
The component 
\be
\psi_\pm(\pm p)^{0\dots 01\dots
1}_{A_1\dots A_rA'_{r+1}\dots A'_n}
\ee
of 
\be
\psi_\pm(\pm p)_{\alpha_1\dots \alpha_n}=
\chi{^{(+)}_{\pm\alpha_1}}\dots \chi{^{(+)}_{\pm\alpha_n}}
f_\pm(\pm p)^{(+\dots +)}+\dots +
\chi{^{(-)}_{\pm\alpha_1}}\dots \chi{^{(-)}_{\pm\alpha_n}}
f_\pm(\pm p)^{(-\dots -)}
\ee
contains only one expression involving only $\pi_A$'s or $\bar
\pi_{A'}$'s, namely
\be
(-1)^nN(\omega,\pi)^{n}
\pi_{A_1}\dots \pi_{A_r}
\bar \pi_{A'_{r+1}}\dots \bar \pi_{A'_n}
f_\pm(\pm p)\underbrace{^{(-\dots -}}_r
\underbrace{^{+\dots +)}}_{n-r}.
\ee
All other parts of the expansion contain at least one $\omega_A$
or $\bar \omega_{A'}$ and hence are annihilated if contracted
with 
$$
\omega^{A_1}\dots \omega^{A_r}
\bar \omega^{A'_{r+1}}\dots \bar \omega^{A'_n}.
$$
Therefore 
\be
N(\omega,\pi)^{-n}
\omega^{A_1}\dots \omega^{A_r}
\bar \omega^{A'_{r+1}}\dots \bar \omega^{A'_n}
\psi_\pm(\pm p)^{0\dots 01\dots
1}_{A_1\dots A_rA'_{r+1}\dots A'_n}
=f_\pm(\pm p)\underbrace{^{(-\dots -}}_r
\underbrace{^{+\dots +)}}_{n-r}=
f_\pm(\pm p)^{0\dots 01\dots 1}.
\ee
In the last formula we have changed the convention analogously
to this from Appendix~\ref{APL}.

\subsection{Generalized Gross-Kaiser norm: An alternative covariant
proof of invariance}
\label{A3}

The other form of potentials 
is not very helpful in proving invariance of the Bargmann-Wigner
norm in the general spin case. It is instructive, however, to see 
how the spinor language simplifies the standard
proof in the particular case of the Maxwell field (cf. \cite{K}
and \cite{G}). 

Consider the electromagnetic spinor 
\be
\varphi_{\pm}(\pm p)_{AB}=\frac{i}{2}F_\pm^{qr}(\pm p)
\sigma_{qrAB},
\ee
which satisfies 
\be
\varphi_{\pm}(\pm p)_{AB}=\mp ip_{AA'}\phi_\pm(\pm p){_{B}}{^{A'}}
=\mp ip_{BA'}\phi_\pm(\pm p){_{A}}{^{A'}}
\ee
implying the Lorenz gauge
\be
p_{AA'}\phi_\pm(\pm p){^{AA'}}=0
\ee
for the 4-vector potential $\phi_\pm^a(\pm p)$. 
We consider the tensor
\be
T_{\pm ab}(\pm p)&=&
\varphi_\pm(\pm p)_{AB}\bar \varphi_\pm(\pm p)_{A'B'}=
p_{AC'}\phi_\pm(\pm p){_{B}}{^{C'}}
p_{CA'}\phi_\pm(\pm p){^{C}}{_{B'}}
=
p_{AA'}p_{CC'}\phi_\pm(\pm p){_{B}}{^{C'}}
\phi_\pm(\pm p){^{C}}{_{B'}}\nonumber\\
&=&
p_{AA'}p_{BC'}\phi_\pm(\pm p){_{C}}{^{C'}}
\phi_\pm(\pm p){^{C}}{_{B'}}=
-\frac{1}{2}p_{a}p_{b}\phi{_{\pm c}}(\pm p)
\phi{_{\pm}^c}(\pm p),
\ee
where we have used the trace-reversal identity and the fact that
\be
\phi_\pm(\pm p){_{CC'}}
\phi_\pm(\pm p){^{C}}{_{B'}}=
-\phi_\pm(\pm p){_{CB'}}
\phi_\pm(\pm p){^{C}}{_{C'}}.
\ee
The tensor satisfies the formula
\be
T_{\pm ab}(\pm p)&=&
\frac{1}{2}\Bigl(
\frac{1}{4}g_{ab}F_{\pm cd}(\pm p)F_\pm^{cd}(\pm p)
-F_{\pm ac}(\pm p)F{_{\pm b}}^{c}(\pm p)\Bigr),
\ee
and, in particular, 
\be
T_{\pm 00}(\pm p)&=&
\frac{1}{4}\Bigl(
\bbox E_\pm(\pm p)^2 + \bbox B_\pm(\pm p)^2\Bigr),
\ee
where $\bbox E_\pm(\pm p)$ and $\bbox B_\pm(\pm p)$ are the
positive and negative frequency Fourier transforms of the
electromagnetic field. 

Now we can repeat the reasoning presented above for the general
case and the norm used in the wavelet analysis of the
electromagnetic field \cite{K} becomes a particular case of
\be
\parallel \varphi\parallel^2=
\parallel \varphi_+\parallel'^2
+
\parallel\varphi_-\parallel'^2,
\ee
where
\be
\parallel \varphi_\pm\parallel'^2=
\int\frac{d^3p}{2p^0}
\frac{
t_1^{a}t_2^{b}
T_{\pm ab}(\pm p)
}{
t_1^{c} t_2^{d}p_{c} p_{d}}.
\ee

\end{document}